\documentclass{emulateapj}
\newcommand{\degree}{\hbox{$^\circ$}}
\newcommand{\dgr}{$\cal{D}$}
\newcommand{\etal}{et\,al.}
\newcommand{\halpha}{H$\alpha$}
\newcommand{\gsim}{\raise0.3ex\hbox{$>$}\kern-0.75em{\lower0.65ex\hbox{$\sim$}}}

\newcommand{\lsim}{\raise0.3ex\hbox{$<$}\kern-0.75em{\lower0.65ex\hbox{$\sim$}}}

\newcommand{\msun}{M$_{\odot}$}
\newcommand{\mm}{$\mu$m}
\newcommand{\pom}{\,$\pm$\,}
\newcommand{\sings}{{\it SINGS}}
\newcommand{\snu}{S$_{\nu}$}
\newcommand{\spitzer}{{\it Spitzer}}

\newcommand{\HI}{H~{\sc i}}
\newcommand{\HII}{H~{\sc ii}}
 
\begin{document}     
\slugcomment{Accepted for publication in the Astrophysical Journal}
\title{Warm Dust and Spatially Variable PAH Emission\\ in the Dwarf Starburst
Galaxy NGC 1705}

\author{John M. Cannon\altaffilmark{1,2},
John-David T. Smith\altaffilmark{3},
Fabian Walter\altaffilmark{1},
George J. Bendo\altaffilmark{4,3},
Daniela Calzetti\altaffilmark{5},
Daniel A. Dale\altaffilmark{6},
Bruce T. Draine\altaffilmark{7},
Charles W. Engelbracht\altaffilmark{3},
Karl D. Gordon\altaffilmark{3},
George Helou\altaffilmark{8}
Robert C. Kennicutt, Jr.\altaffilmark{9,3},
Claus Leitherer\altaffilmark{5},
Lee Armus\altaffilmark{8},
Brent A. Buckalew\altaffilmark{8},
David J. Hollenbach\altaffilmark{10},
Thomas H. Jarrett\altaffilmark{8},
Aigen Li\altaffilmark{11},
Martin J. Meyer\altaffilmark{5},
Eric J. Murphy\altaffilmark{12},
Michael W. Regan\altaffilmark{5},
George H. Rieke\altaffilmark{3},
Marcia J. Rieke\altaffilmark{3},
H{\'e}l{\`e}ne Roussel\altaffilmark{1},
Kartik Sheth\altaffilmark{8},
Michele D. Thornley\altaffilmark{13}}

\altaffiltext{1}{Max-Planck-Institut f{\"u}r Astronomie, K{\"o}nigstuhl 17,
D-69117, Heidelberg, Germany; walter@mpia.de, roussel@mpia.de}
\altaffiltext{2}{Current address: Astronomy Department, Wesleyan University, 
Middletown, CT 06459; cannon@astro.wesleyan.edu}
\altaffiltext{3}{Steward Observatory, University of Arizona, 933 North Cherry
Avenue, Tucson, AZ 85721; gbendo@as.arizona.edu, cengelbracht@as.arizona.edu,
kgordon@as.arizona.edu, robk@as.arizona.edu, grieke@as.arizona.edu, 
mrieke@as.arizona.edu, jdsmith@as.arizona.edu}
\altaffiltext{4}{Astrophysics Group, Imperial College, Blackett Laboratory,
Prince Consort Road, London SW7 2AZ United Kingdom; g.bendo@imperial.ac.uk}
\altaffiltext{5}{Space Telescope Science Institute, 3700 San Martin Drive,
Baltimore, MD 21218; calzetti@stsci.edu, leitherer@stsci.edu, 
martinm@stsci.edu, mregan@stsci.edu}
\altaffiltext{6}{Department of Physics and Astronomy, University of Wyoming,
Laramie, WY 82071; ddale@uwyo.edu}
\altaffiltext{7}{Princeton University Observatory, Peyton Hall, Princeton, NJ
08544; draine@astro.princeton.edu}
\altaffiltext{8}{California Institute of Technology, MC 314-6, Pasadena, CA
91101; gxh@ipac.caltech.edu, lee@ipac.caltech.edu, brentb@ipac.caltech.edu,
jarrett@ipac.caltech.edu, kartik@astro.caltech.edu}
\altaffiltext{9}{Institute of Astronomy, University of Cambridge, Madingley 
Road, Cambridge CB3 0HA, UK; robk@ast.cam.ac.uk}
\altaffiltext{10}{NASA/Ames Research Center, MS 245-6, Moffett Field, CA,
94035; hollenba@ism.arc.nasa.gov}
\altaffiltext{11}{Department of Physics and Astronomy, University of
Missouri, Columbia, MO 65211; lia@missouri.edu}
\altaffiltext{12}{Department of Astronomy, Yale University, New Haven, CT 
06520; murphy@astro.yale.edu}
\altaffiltext{13}{Department of Physics, Bucknell University, Lewisburg, PA
17837; mthornle@bucknell.edu}

\begin{abstract}

We present {\it Spitzer} observations of the nearby dwarf starburst galaxy
NGC\,1705 obtained as part of the {\it Spitzer Infrared Nearby Galaxies
Survey}. The galaxy morphology is very different shortward and longward of
$\sim$ 5 \mm: optical and short-wavelength IRAC imaging shows an underlying
red stellar population, with the central super star cluster (SSC) dominating
the luminosity; longer-wavelength IRAC and MIPS imaging reveals warm dust
emission arising from two off-nuclear regions that are offset by $\sim$ 250 pc
from the SSC and that dominate the far-IR flux of the system.  These regions
show little extinction at optical wavelengths. The galaxy has a relatively low
global dust mass ($\sim$ 2\,$\times$\,10$^{5}$ \msun, implying a global
dust-to-gas mass ratio $\sim$ 2--4 times lower than the Milky Way average,
roughly consistent with the metallicity decrease). The off-nuclear dust
emission appears to be powered by photons from the same stellar population
responsible for the excitation of the observed \halpha\ emission; these
photons are unassociated with the SSC (though a contribution from embedded
sources to the IR luminosity of the off-nuclear regions cannot be ruled
out). Low-resolution IRS spectroscopy shows moderate-strength PAH emission in
the 11.3 \mm\ band in the more luminous eastern peak; no PAH emission is
detected in the SSC or the western dust emission complex. There is significant
diffuse emission in the IRAC 8 \mm\ band after starlight has been removed by
scaling shorter wavelength data; the fact that IRS spectroscopy shows
spatially variable PAH emission strengths compared to the local continuum
within this diffuse gas suggests caution in the interpretation of IRAC diffuse
8 \mm\ emission as arising from PAH carriers alone.  The nebular metallicity
of NGC\,1705 falls at the transition level of $\sim$ 0.35 Z$_{\odot}$ found by
Engelbracht and collaborators, below which PAH emission is difficult to
detect; the fact that a system at this metallicity shows spatially variable
PAH emission demonstrates the complexity of interpreting diffuse 8 \mm\
emission in galaxies. NGC\,1705 deviates significantly from the canonical
far-infrared vs. radio correlation, having significant far-infrared emission
but no detected radio continuum.

\end{abstract}						


\section{Introduction}
\label{S1}

Local dwarf galaxies serve as potential examples of young galaxies in
formation. Due in part to their typically low metal abundances, the
characteristics of the interstellar medium (ISM) differ markedly from those in
larger, more metal-rich systems. In cases where spatially and temporally
concentrated star formation occurs, ``dwarf starburst'' galaxies inject
energies into the ISM that can be a substantial fraction of their system
binding energies; these bursts overwhelm the shallow potential wells of the
parent galaxy and result in spectacular galactic winds
\citep[e.g.,][]{heckman01a,martin02,ott03}.  Starbursts in low-mass dwarfs
represent the low-luminosity end of the starburst superwind spectrum; outflows
increase in strength up to the ultraluminous infrared galaxies
\citep{heckman90}.

In extreme cases, single star clusters may input sufficient kinetic
energy to drive galactic winds in dwarf galaxies.  The responsible
``super-star clusters''(SSCs) have typical ages $\lsim$ 1 Gyr and
dynamical masses $\sim$ 10$^5$--10$^8$ \msun; the most massive may be
globular cluster progenitors.  While many massive starburst and
interacting systems are able to attain the densities and pressures
requisite for SSC formation (i.e., n$_{\rm H} \sim$ 10$^6$ cm$^{-3}$,
P/k $\gsim$ 10$^8$--10$^9$ cm$^{-3}$ K, radii $\lsim$ 5 pc; see
{O'Connell 2004}\nocite{oconnell04} for a recent review), the
formation rate and maximum luminosity of SSCs appear to generally
scale with the total star formation rate; typically these clusters
form a power-law mass distribution.  Hence, massive, interacting
galaxies such as the ``Antennae'' (NGC\,4038/4039) are able to produce
hundreds of massive young clusters \citep{whitmore95}.  Dwarf
galaxies, on the other hand, produce relatively few massive SSCs, and
their formation may follow a different track than those of massive
clusters in larger galaxies; the cluster luminosity functions appear
to be discontinuous in dwarfs, with SSCs being much more massive than
other clusters (though this may be a statistical effect; see {Whitmore
2004}\nocite{whitmore04conf}).

At a distance of 5.1\,$\pm$\,0.6 Mpc \citep{tosi01}, NGC\,1705 is one of the 
best-studied dwarf starburst galaxies.  The optical emission is dominated by 
the ``super-star cluster'' (SSC) NGC\,1705-1 \citep{meurer95}, a compact
\citep[effective radius $\sim$ 1.5 pc;][]{smith01}, massive \citep[dynamical
mass $\sim$ 10$^5$ \msun;][]{ho96b} cluster of age $\lsim$ 12 Myr 
\citep{vazquez04}.  The age of the cluster and the amount of ionized gas
relative to the UV luminosity are consistent with most of the O stars having 
died out, so the system may be more correctly deemed a ``post-starburst''
galaxy \citep{heckman97}. Dominant SSCs are comparatively rare in dwarf 
starburst galaxies, and NGC\,1705 is one of the few systems where the 
luminosity of a single cluster dominates the galaxy's UV light.  The recent 
star formation is driving a multi-phase outflow from the disk, seen in the 
neutral, ionized, coronal, and X-ray emitting gas phases 
\citep{heckman97,hensler98,heckman01a}.  The \HI\ dynamics show
normal solid-body rotation, though extra-planar gas may also be involved with 
the outflow \citep{meurer98}.

NGC\,1705 has undergone strong recent star formation; \citet{annibali03}
compare synthetic color-magnitude diagrams to optical and near-infrared
photometry to estimate a rate of 0.3 \msun\,yr$^{-1}$; combining with the
total \HI\ mass of 1.04\,$\times$\,10$^8$ \msun\ (using the integrated flux
from {Meurer \etal\ 1998}\nocite{meurer98} scaled to our adopted distance)
implies a gas depletion timescale of only $\sim$ 0.35 Gyr. Given the galaxy's
recent starburst, its sub-solar nebular metallicity [Z $\sim$ 35\% Z$_{\odot}$
(using the solar oxygen abundance of {Asplund \etal\ 2004}\nocite{asplund04});
{Lee \& Skillman 2004}\nocite{lee04a}], its gas-rich central disk (N$_{\rm H}$
$>$ 10$^{21}$ cm$^{-2}$; {Meurer \etal\ 1998}\nocite{meurer98}), and the
presence of a SSC, it presents an ideal example of a starbursting dwarf galaxy
with an extreme star-forming ISM.  We present \spitzer\ imaging and
spectroscopy of this system, obtained as part of the {\it Spitzer Infrared
Nearby Galaxies Survey} \citep[\sings;][]{kennicutt03}, with the aim of
characterizing the nature of warm dust and PAH emission in the metal-poor,
active ISM.  This system has one of the highest current star formation rates
of any dwarfs in the \sings\ sample and is the only dwarf in the survey that
contains a known SSC.

\begin{figure*}
\begin{center}
\includegraphics[width=12 cm]{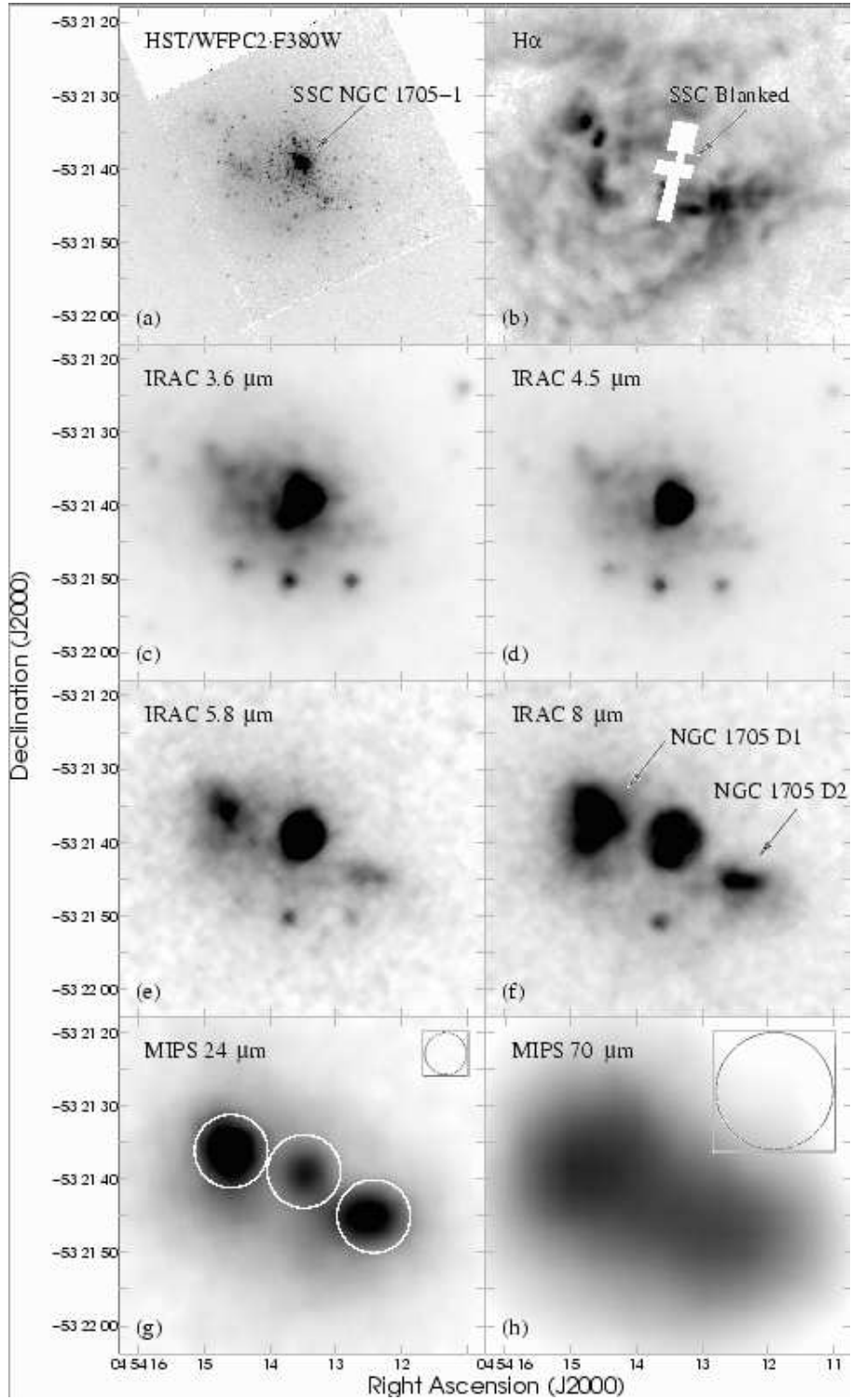}
\end{center}
\caption{Images of NGC\,1705 at 8 different wavelengths: {\it HST}/WFPC2 F380W
image (a, observed for program GO-7506, P.I. M. Tosi); continuum-subtracted
\halpha\ image from the survey of Gil de Paz \etal\ (2003, b; note that the
SSC and some surrounding regions have been blanked - compare with
Figure~\ref{figcap5}b); IRAC 3.6, 4.5, 5.8, and 8 \mm\ (c, d, e, f); MIPS 24
and 70 \mm\ (g, h). The white circles in (g) denote the apertures used to
extract flux densities at 10\arcsec\ resolution (see text and Table~\ref{t1});
FWHM sizes of the MIPS bands are shown as boxed circles in (g) and (h).  Note
the striking morphological differences between the dust and stellar continuum
longward and shortward of $\sim$ 5 \mm.}
\label{figcap1}
\end{figure*}

\section{Observations and Data Reduction}
\label{S2}

For a complete overview of the \sings\ observational strategies, see
\citet{kennicutt03}.  NGC\,1705 was observed in standard IRAC imaging
mode (producing two 5.2\arcmin\,$\times$\,5.2\arcmin\ fields, offset
by $\sim$ 1\arcmin) and in MIPS scan mapping mode (covering
$\sim$0.5\degree, with foreground and background sources dominating the
emission at $\gsim$1.5\,$\times$\,D$_{25}$ and beyond). IRS spectral
mapping was performed using all four modules; the resulting maps cover
an area larger than the optical body in all modules.  IRS and IRAC
data were obtained in 2004 July (62.4 minutes in IRS long-low and
short-low modules, and 18.5 minutes in IRAC imaging); MIPS data were
obtained in 2004 September (58.6 minutes in all three modules).

All data were processed by the \sings\ pipelines.  The IRAC pipeline
reduces basic calibrated data images; flux levels are uncertain at the
$\sim$ 10\% level due to systematics (see discussion of aperture
correction effects in \S~\ref{S3}).  The MIPS Instrument Team Data
Analysis Tool \citep{gordon05} was used to process the MIPS data,
while the {\it Cubism} tool (J.~D. Smith \etal, in preparation) was
used for spectral extraction and analysis of the IRS data.  Systematic
uncertainties limit the absolute flux calibration to $\sim$ 10\% in
the MIPS 24 \mm\ band and to $\sim$ 20\% in the MIPS 70 and 160 \mm\
bands; the flux uncertainty of the IRS data is $\sim$ 20\%. The FWHMs
of the MIPS PSFs are 6\arcsec, 18\arcsec, and 38\arcsec\ at 24, 70,
and 160 \mm, respectively. Global fluxes were measured after
foreground/background source removal and background subtraction, using
matched 105\arcsec\ radius circular apertures, centered on the SSC, in
all MIPS and IRAC wavebands; this radius is slightly larger than the
fields of view shown in Figure~\ref{figcap1}.  No emission associated
with NGC\,1705 is present outside the area shown in Figure~\ref{figcap1}.
Since our measured global flux densities agree within the errors with
those published in \citet{dale05c}, we adopt their values as our final
global flux density measurements.  The 11 \mm\ PAH map was created
from the full spectral data cube by averaging the flux density over
the wavelength region 11.115 -- 11.860 \mm, and subtracting the
average continuum in the flanking regions (10.805 -- 11.115 \mm\ and
11.860 -- 12.357 \mm).

The \halpha\ image analyzed here was taken from the public archive of the
Palomar/Las Campanas Imaging Atlas of Blue Compact Dwarf Galaxies
\citep{gildepaz03}.  These images were obtained with the du~Pont 100-inch
telescope of the Las Campanas Observatory.  The total flux of the galaxy was
matched to the value presented in \citet{gildepaz03}; note that the SSC has
been removed from the \halpha\ image used in this paper.

\section{Warm Dust and PAH Emission in NGC\,1705}
\label{S3}

Figure~\ref{figcap1} shows images of NGC\,1705 at 8 different wavelengths.
(a) shows the near-UV ($\lambda \sim$ 380\,nm) continuum, where the SSC is the
brightest single UV source in the galaxy \citep[producing more than $\sim$
40\% of the UV photons;][]{vazquez04}.  There is an underlying stellar
continuum containing a sizeable fraction of intermediate-mass stars: in the
Planetary Camera field of view shown in Figure~\ref{figcap1}(a),
\citet{tosi01} find that $\sim$ 50\% of the detected stars have masses $\gsim$
3 \msun\ (corresponding to an age of $\lsim$ 500 Myr), while $\sim$ 10\% have
masses $\gsim$ 7 \msun\ (ages $\lsim$ 60 Myr). \halpha\ emission is present
throughout the optical body of the system (see Figure~\ref{figcap1}b; note
that the SSC has been blanked in this image); the highest-surface brightness
emission is offset from the SSC by $\sim$ 250 pc, coincident with the
morphology of dust observed in the longer-wavelength \spitzer\ bands (see
below). 

Moving into the near- and mid-infrared (IR), the IRAC 3.6 and 4.5 \mm\ images
(see Figures~\ref{figcap1}c, d) are sensitive to the underlying red stellar
population (typically consisting of young red supergiants and an older stellar
population). The 3.6 \mm\ band is also known to contain some PAH emission
bands \citep{helou00,lu03,helou04} as well as emission from \HII\ gas
processes \citep[e.g.,][]{wang04}.  The IRAC 5.8 and 8 \mm\ bands (see
Figures~\ref{figcap1}e, f) are primarily sensitive to dust and PAH emission,
though a stellar component is present as well \citep[e.g.,][]{pahre04a}. Note
that the SSC retains a high surface brightness in the near- and mid-IR, though
by 24 \mm\ the off-nuclear dust emission regions contribute the bulk of the
global flux density (see Figure~\ref{figcap1}g).

\begin{deluxetable*}{lcccc}[!ht]
\tablecaption{Observed Emission Properties of NGC\,1705\tablenotemark{a}}
\tablewidth{0pt}
\tablehead{
\colhead{Parameter}
&\colhead{NGC 1705 D1}
&\colhead{NGC 1705 D2}
&\colhead{SSC NGC 1705-1}
&\colhead{NGC 1705 Total Galaxy}}
\startdata 
\vspace{0.1 cm}
$\alpha$ (J2000) &04:54:14.59    &04:54:12.42    &04:54:13.48    &04:54:13.5\\
$\delta$ (J2000) &$-$53:21:36.20 &$-$53:21:45.07 &$-$53:21:38.95 &$-$53:21:40\\
\halpha\ Flux\tablenotemark{b} &1.8\pom0.3 &2.2\pom0.3 &0.92\pom0.14
&25.3\pom3.8\\
IRAC 3.5\,\mm\ Flux Density\tablenotemark{a}  &1.8\pom0.2 &0.8\pom0.08 &6.6\pom0.7 &28\pom3\\
IRAC 4.5\,\mm\ Flux Density\tablenotemark{a}  &1.2\pom0.2 &0.7\pom0.07 &4.6\pom0.5 &19\pom2\\
IRAC 5.7\,\mm\ Flux Density\tablenotemark{a}  &1.9\pom0.2 &0.9\pom0.09 &4.1\pom0.5 &12\pom2\\
IRAC 7.9\,\mm\ Flux Density\tablenotemark{a}  &3.4\pom0.4 &1.7\pom0.2  &4.5\pom0.5 &22\pom2\\
MIPS 24\,\mm\ Flux  Density\tablenotemark{a}  &6.0\pom0.6 &4.8\pom0.5  &3.9\pom0.4 &52\pom5\\
MIPS 70\,\mm\ Flux  Density\tablenotemark{a}  &126\pom63\tablenotemark{c}
&109\pom55\tablenotemark{c} &N/A    &1090\pom220\\
MIPS 160\,\mm\ Flux Density\tablenotemark{a}  &N/A    &N/A    &N/A    &1200\pom250\\
\enddata
\enddata
\label{t1}\vspace{-0.4 cm}
\tablenotetext{a}{All measurements are in units of mJy, unless
otherwise noted; \spitzer\ flux densities derived without aperture
corrections.  Apertures of radius 5.7\arcsec\ are used to measure the
flux densities presented, unless otherwise noted.}\\
\tablenotetext{b}{Derived using the \halpha\ image presented in
\citet{gildepaz03}, assuming their global \halpha\ flux; units of
10$^{-13}$ erg\,s$^{-1}$\,cm$^{-2}$.}  
\tablenotetext{c}{Fluxes of NGC\,1705 D1 and D2 at 70 \mm\ are
measured using apertures of 9\arcsec\ radius (i.e., the FWHM of the 70
\mm\ beam).  Errorbars reflect both systematics and the potential
contribution from the SSC (estimated by assuming a flat SED as
suggested from the data between 4.5 and 24 \mm).}
\end{deluxetable*}

There is a striking difference in the galaxy's morphology shortward and
longward of $\sim$ 5 \mm.  There is a negligible amount of off-nuclear dust
emission in the 4.5 \mm\ band, while longer wavelengths show two strong mid-
and far-IR emission peaks that are offset from the SSC and coincident with the
\halpha\ surface brightness maxima. These dust emission regions, labeled as
NGC\,1705 D1 and D2 (see Figure~\ref{figcap1}f), are the highest-surface
brightness regions in NGC\,1705 in the MIPS bands (together contributing
$\sim$ 25\% of the 24 \mm\ global flux density; see below).  We discuss the
nature of the off-nuclear emission in more detail in \S~\ref{S3.1}.

We measure the flux densities of the three individual dust emission peaks
(NGC\,1705 D1, D2, and NGC\,1705 SSC-1), at a resolution slightly larger than
that of the MIPS 24 \mm\ band (FWHM $=$5.7\arcsec), in the IRAC, MIPS 24 \mm,
and \halpha\ images.  Fluxes were extracted using non-overlapping apertures of
10\arcsec\ diameter, centered on the locations given in Table~\ref{t1} and
shown in Figure~\ref{figcap1}.  Due to the distance and size of NGC\,1705, we
lose sufficient resolution by the 70 \mm\ band to distinguish the individual
emission peaks.  No aperture corrections were applied to the measured flux
densities; aperture corrections for extended sources are estimated to be
between $\sim$ 4--25\% \citep[e.g., ][]{pahre04a}.  For point sources, the
aperture corrections are estimated to be of order 5--7\% in the IRAC bands for
this aperture size, while at 24 \mm\ the correction may be as large as a
factor of 2; this effect may be important for the discussion in \S~\ref{S3.1}
and \ref{S3.2}.  The values listed in Table~\ref{t1} demonstrate that the
spectral energy distributions (SEDs) from both NGC\,1705 D1 and D2 are
beginning to rise towards a far-IR peak characteristic of dust emission in
normal galaxies \citep[see, e.g.,][]{dale05c}.  The SED of the SSC, however,
appears to flatten out by the MIPS 24 \mm\ band, suggesting that it
contributes only a small fraction of the total far-IR luminosity of the
system; SED model fits to each of the emission peaks is not feasible, given
the lack of resolution at the longer MIPS wavelengths (but see {Dale \etal\
2005}\nocite{dale05c} and \S~\ref{S3.1} for the galaxy's global SED and model
fit).

\subsection{The Origin and Characteristics of Nebular and Dust Emission at D1 
and D2}
\label{S3.1}

\begin{figure*}
\plotone{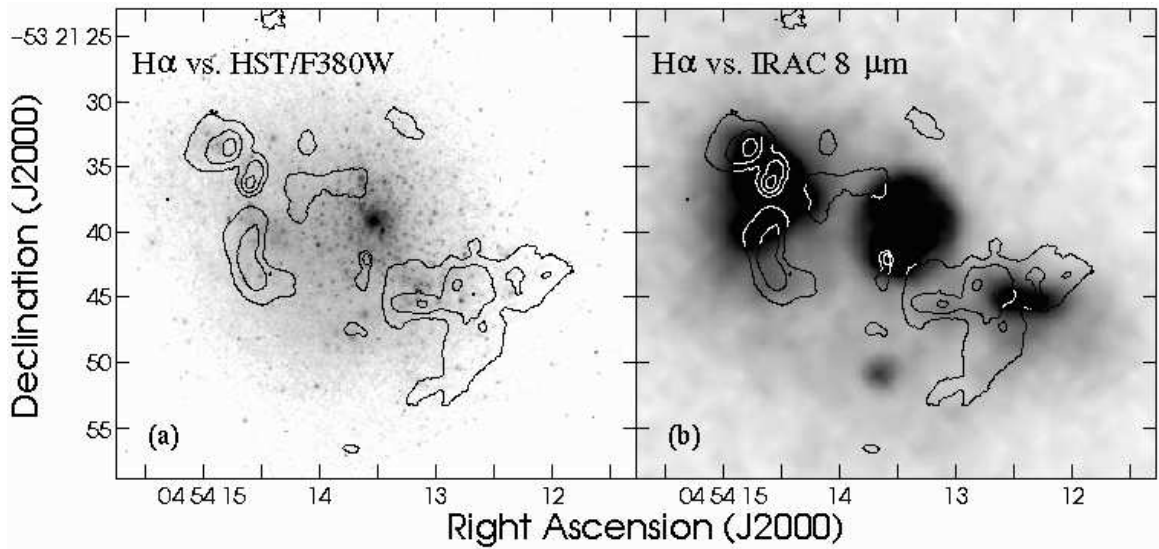}
\caption{Closer view of the optical body of NGC\,1705, overlaid with
contours of the high-surface brightness \halpha\ emission. (a) shows
the {\it HST}/WFPC2 image displayed in a log stretch to compress the
dynamic range and highlight field stars and clusters, while (b) shows
the IRAC 8 \mm\ image.  Contours are at \halpha\ flux levels of (1.3,
2.6, 5.2)\,$\times$\,10$^{-16}$ erg\,s$^{-1}$\,cm$^{-2}$.  Note that
there are massive field stars associated with the regions of \halpha\
and dust emission.}
\label{figcap2}
\end{figure*}

As shown in Figures~\ref{figcap1} and \ref{figcap2}, both the IR and
\halpha\ emission peaks (NGC\,1705 D1 and D2) are displaced from the
SSC by $\gsim$ 250 pc.  \citet{vazquez04} use UV spectroscopy to
demonstrate that the SSC produces the bulk of the UV luminosity of
NGC\,1705, but too few ionizing photons to produce the \halpha\
emission seen hundreds of pc away at the locations of D1 and D2.
Thus, young, massive stars (stellar masses in the range 10-30 \msun,
with corresponding main sequence lifetimes $\simeq$ 5--30 Myr,
depending on stellar evolution models and adopted metallicity; see
{Tosi \etal\ 2001}\nocite{tosi01} and {Annibali \etal\
2003}\nocite{annibali03}) appear to be the source of the ionizing
photons leading to the observed off-nuclear \halpha\ emission.

To assess whether the stellar population creating this \halpha\ emission also
powers the observed dust emission at locations D1 and D2, it is instructive to
compare the star formation rates derived from \halpha\ and from the IR.  Due
to the distance of the system, the physical resolution at 70 and 160 \mm\
precludes the use of a total IR-based star formation rate for the individual
peaks D1 and D2 \citep[e.g., ][]{dale02}. We instead apply the monochromatic
calibration between \halpha\ and 24 \mm\ luminosity derived in M\,51 by
\citet{calzetti05}:
\begin{equation}     
log(L_{24}) = 1.03log({L_{H\alpha}}) - 0.06945
\label{eq1}
\end{equation}
where L$_{\rm 24}$ and L$_{\rm H\alpha}$ are the observed luminosities at 24
\mm\ and \halpha, respectively; the intrinsic \halpha/P$\alpha$ ratio is
assumed to be 8.734; and we apply the central wavelength of 24.0 \mm\ to
convert the 24 \mm\ flux density to a luminosity measurement.  Performing this
comparison, we find that the observed 24 \mm\ luminosities of D1 and D2 are
factors of 3 and 5 lower than predicted from the \halpha\ luminosities
alone. Note that extinction at \halpha, while minimal [\citet{lee04a} find
only marginal A$_{\rm V}$ values of 0.0 and 0.47 mag for D1 and D2,
respectively], would make this discrepancy larger; aperture corrections at 24
\mm\ (which may be as large as a factor of two) will bring these values into
better agreement.

A lower star formation rate derived from the IR compared to \halpha\ has been
found for other dwarf galaxies in the \sings\ sample ({Cannon \etal\
2005}\nocite{cannon05}; Walter \etal, in prep.).  This effect has been
attributed, to first order, to the lower dust contents found in the metal-poor
ISM.  In NGC\,1705, \halpha\ and 24 \mm\ star formation rates agree within a
factor of a few, suggesting that a substantial component of the dust emission
at D1 and D2 is powered by re-radiation of photons from the same sources that
are responsible for the \halpha\ emission.  The offset may be due to
differences in ISM physics between metal-rich and metal-poor galaxies.  While
the energetics are comparable, we cannot rule out a contribution from embedded
sources that do not contribute to the observed \halpha\ emission.  In this
regard, it is worth noting that some low-metallicity galaxies harbor
heavily-embedded star formation regions (e.g., SBS\,0335$-$052; {Houck \etal\
2004}\nocite{houck04}, {Hunt \etal\ 2004}\nocite{hunt04}).

\begin{figure*}
\plotone{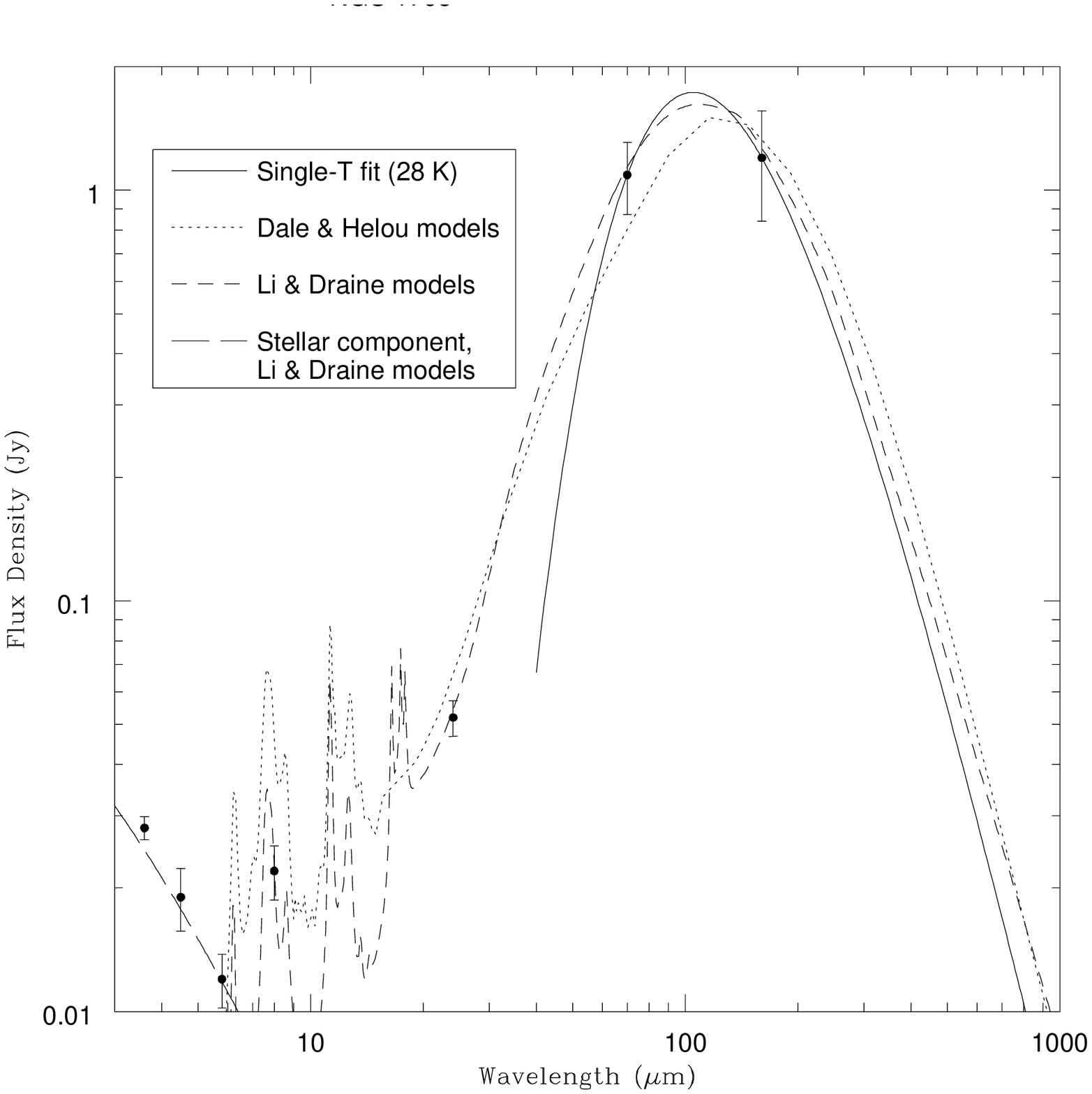}
\caption{Three independent fits to the observed flux densities in NGC\,1705.
The solid line shows a single-temperature fit to the 70 and 160 \mm\ data
points; note that it strongly underestimates the observed 24 \mm\ flux
density.  A single-temperature blackbody is thus unable to simultaneously fit
the 24, 70 and 160 \mm\ data points.  The dotted line shows a fit of the Dale
\& Helou (2002) SED models to the 24, 70 and 160 \mm\ data, which improves the
fit at short wavelengths compared to the single-temperature fit.  The
short-dashed line shows a fit of the {Li \& Draine} models to the
observations; this model suite is best able to reproduce the observed flux
densities.  The long-dash line shows the stellar component output from the Li
\& Draine (2001, 2002) model fit.  Given the uncertainties in the
long-wavelength MIPS data ($\sim$ 20\%), we consider both the Dale \& Helou
(2002) and the Li \& Draine (2001, 2002) models to provide satisfactory fits
to the observations; we adopt the mean of the dust masses predicted by these
two model suites as our final estimate of the dust mass in NGC\,1705 (see
discussion in text).}
\label{figcap3}
\end{figure*}

An estimate of the dust mass in each component is not possible, given
our inability to resolve the two peaks at 160 \mm.  The total
dust mass in NGC\,1705 can be very simplistically estimated by
applying the following relation:
\begin{equation}     
M_{dust} = \frac{D^2 f_{\nu}}{\kappa_{\nu} B(T)}
\label{eq2}
\end{equation}
where D is the distance to the galaxy, $f_{\nu}$ is the flux density,
$\kappa_{\nu}$ represents the absorption opacity of the dust at
frequency $\nu$ (given in Li \& Draine 2001\nocite{li01}), and $B(T)$
is the Planck function evaluated at temperature T, derived by fitting
a blackbody modified by a $\lambda^{-2}$ emissivity function to the 70
and 160 \mm\ data. As shown in Figure~\ref{figcap3}, such a
single-temperature solution is unable to simultaneously fit the
observed 24, 70 and 160 \mm\ data; the fit to the 70 and 160 \mm\ flux
densities grossly underestimates the observed 24 \mm\ flux density,
implying a two-component dust grain population (e.g., cool grains at
$\sim$ 28 K, warmer grains at $\sim$ 55 K).  Similar conclusions are
reached based on {\it ISO} observations of nearby galaxies
\citep{haas98,popescu02,hippelein03}.

The inability of a single blackbody function to fit the observed flux
densities suggests that a more sophisticated model of the far-IR SED
is needed.  The SED models of \citet{dale01} and \citet{dale02}
provide a more robust treatment of the multiple grain populations that
contribute to the far-IR emission in a galaxy (see {Dale \etal\
2005}\nocite{dale05c} for an application of this method to the entire
\sings\ sample).  Using the fit parameters for the global SED of
NGC\,1705 presented in \citet{dale05c} and shown in
Figure~\ref{figcap3}, the models of \citet{dale02} imply that the
dust mass derived using single-temperature fitting (here, to the
observed 70 and 160 \mm\ data) underestimates the true dust mass by a
factor of $\sim$ 9.  This application of the \citet{dale02} models
results in a global dust mass estimate of
(3.8\,\pom\,1.9)\,$\times$\,10$^5$ \msun.

A third estimate of the dust mass in the galaxy can be found using the
models of {Li \& Draine (2001, 2002)}\nocite{li01,li02}.  Here,
radiation field strengths are varied via power-law distributions; PAH,
silicate and graphite grains are illuminated and the resulting SEDs
can be compared to the observations. The fit of these models to our
observations is also shown in Figure~\ref{figcap3}.  The observed SED
is well-reproduced by $\sim$ 2.3\,$\times$\,10$^7$ \msun\ of gas and
dust, assuming the latter is of ``LMC2-type'' (see {Weingartner \&
Draine 2001}\nocite{weingartner01}).  The dust-to-gas ratio (\dgr) of
LMC2-type dust is estimated to be $\sim$ 0.003, which implies a total
dust mass of $\sim$ 7$\times10^4$ \msun\ (with an uncertainty of
$\sim$ 50\%) in NGC\,1705.

Given the uncertainties, we consider the agreement of the dust mass
estimates using the SED models of \citet{dale01} and \citet{dale02}
and the models of \citet{li01,li02} to be satisfactory.  We adopt the
mean value of these two estimates, (2.2\pom1.1)\,$\times$\,10$^5$
\msun, as the global dust mass of NGC\,1705.  Comparing this total
dust mass with the \HI\ mass measured in a matching aperture
(8.9\,$\times$\,10$^7$ \msun; {Meurer \etal\ 1998}\nocite{meurer98})
implies a global \dgr\ of $\sim$ 2.5\,$\times$\,10$^{-3}$, uncertain
by $\sim$ 50\%. This value is a factor of $\sim$ 2--4 lower than the
Galactic average values (\dgr$_{\rm MW}$ $\sim$ 0.006--0.01; {Sodroski
\etal\ 1997}\nocite{sodroski97}, {Li 2005}\nocite{li05conf}), but
larger than values found for the more metal-poor Small Magellanic
Cloud (Z $\sim$ 23\% Z$_{\odot}$; {Dufour 1984}\nocite{dufour84};
average dust-to-gas ratios $\sim$ 10-30 times $<$ \dgr$_{\rm MW}$;
{Stanimirovic \etal\ 2000}\nocite{stanimirovic00}; {Bot \etal\
2004}\nocite{bot04}, {Stanimirovic \etal\
2005}\nocite{stanimirovic05}). If there is an additional amount of
cold dust in NGC\,1705 that radiates longward of 160 \mm\ (similar to
what is seen in some dwarf galaxies with very cold SEDs; e.g., Tuffs
\etal\ 2002), then this value will be a lower limit.

\subsection{Quantifying PAH Emission Strengths}
\label{S3.2}

\begin{figure*}
\plotone{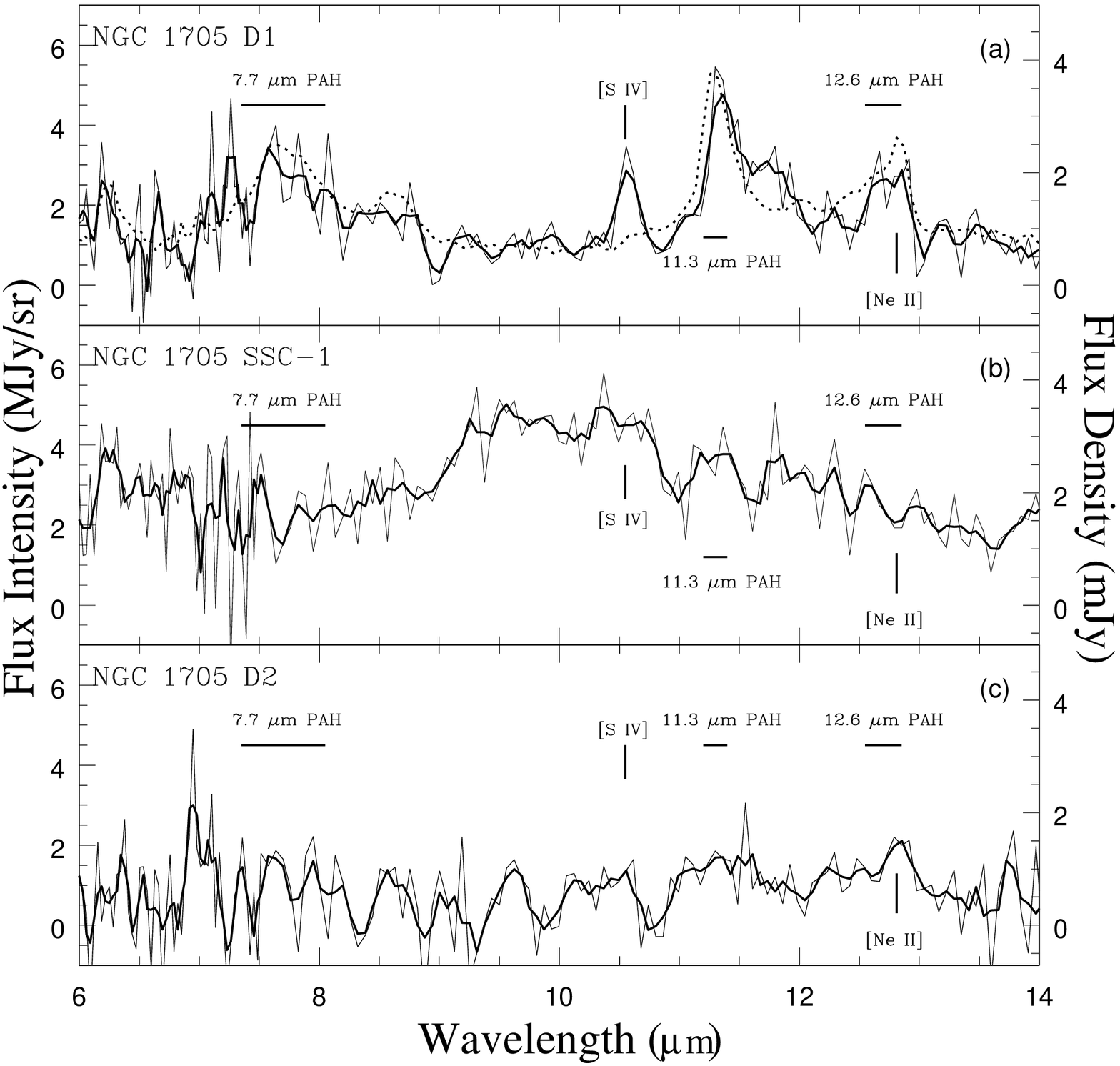}
\caption{IRS spectral extractions in three emission regions of
NGC\,1705, obtained using the IRS Short-Low module; the thin line is
the spectrum at full resolution, while the thick line has been binned
by a median boxcar of 2 pixels to increase the S/N. The extraction
areas are 5.5\arcsec\,$\times$\,5.5\arcsec, centered on the emission
features D1 (panel a), D2 (panel c) and the SSC (panel b; see also
Table~\ref{t1} and Figure~\ref{figcap1}).  In the top panel, the
dotted line shows a representative PAH-dominated spectrum from IRS
observations of the star-forming spiral galaxy NGC\,7331 (see {Smith
\etal\ 2004}), scaled to match the observed flux at the 11.3 \mm\ PAH
feature; note the agreement of the PAH bands with the strengths of the
11.3 and 7.7 \mm\ features observed in NGC\,1705 D1.  We detect the
11.3 \mm\ PAH feature at the $\sim$ 5\,$\sigma$ level in NGC\,1705 D1
only; there is evidence that the continuum also rises toward the 8
\mm\ PAH feature. No PAH emission is associated with NGC\,1705 D2 or
with the SSC.  The 12.6 \mm\ PAH complex is detected at a modest S/N
ratio ($\sim$ 2.5); this feature also has a contribution from [Ne~{\sc
II}] 12.81 \mm\, which is often strong in starburst or AGN
environments.}
\label{figcap4}
\end{figure*}

Since the IRAC 8 \mm\ band samples the 7.7 \mm\ PAH emission complex
as well as the underlying warm dust continuum
\citep[see][]{regan04,engelbracht05}, it is instructive to quantify
the amount of the diffuse emission at 8 \mm\ that is due to PAH
emission. Using the Short-Low IRS module ($\lambda \simeq$ 6-14 \mm, R
$\simeq$ 100), a summation over the wavelength ranges surrounding the
two brightest PAH bands (7.5-9.1 \mm\ and 10.7-11.7 \mm) yields a
detection of the 11.3 \mm\ PAH complex in NGC\,1705 D1 only, with a
nominal rise in the continuum level surrounding the 7.7 \mm\ PAH
feature (though of low S/N; see the spectrum of this region presented
in Figure~\ref{figcap4}). There are no PAH bands detected in NGC\,1705
D2 or towards the SSC. In the top panel of Figure~\ref{figcap4} we
also overlay a typical PAH-dominated spectrum from observations of the
star-forming spiral galaxy NGC\,7331 \citep{smith04}; note the
agreement of the relative strengths of the 11.3 and 7.7 \mm\ PAH bands
in NGC\,1705 D1 with those of the example PAH spectrum.

\begin{figure*}
\plotone{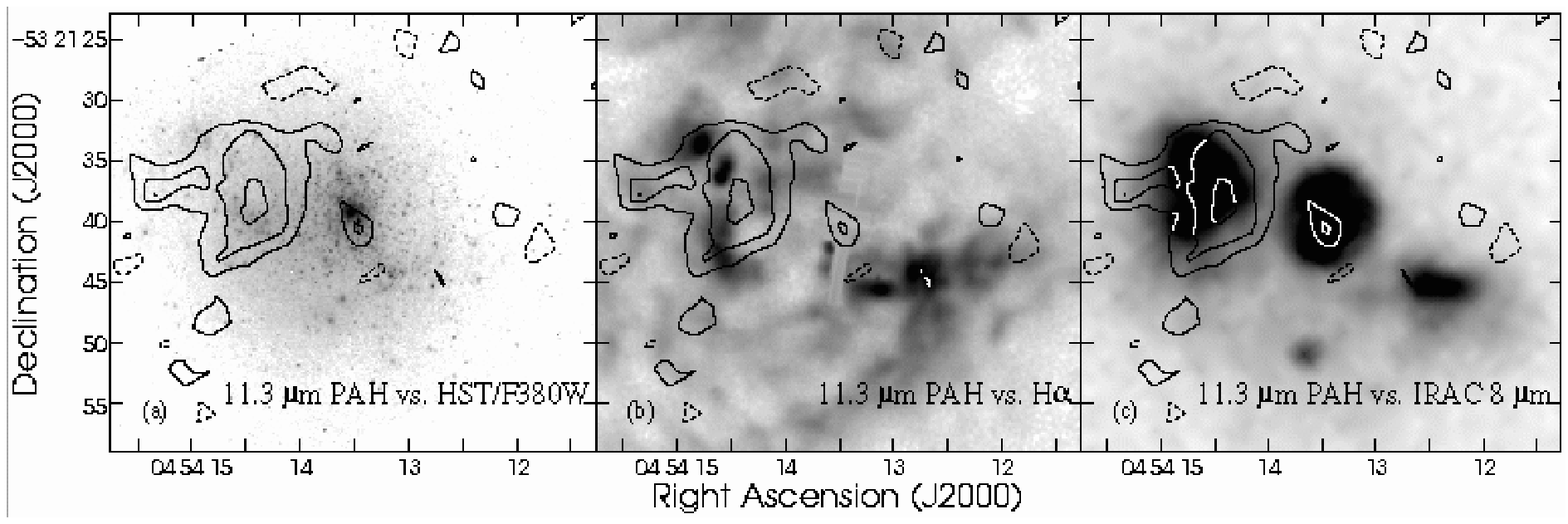}
\caption{Contours of a spectral summation over the IRS short-low data
cube from 10.7 - 11.7 \mm\ (centered on the 11.3 \mm\ PAH feature;
contours at the $\sim$ $-$1.5, 1.5, 3, 4.5\,$\sigma$ levels,
corresponding to $-$0.6, 0.6, 1.2, 1.8 MJy\,sr$^{-1}$, respectively),
overlaid on an {\it HST}/WFPC2 F380W image (a), on the
continuum-subtracted \halpha\ image [b, image from {Gil de Paz \etal\
2003}; note that the SSC has been blanked (see Figure~\ref{figcap1}b
for locations)], and on the IRAC 8 \mm\ image.  No clear PAH emission
is associated with the SSC or with the western dust emission complex
NGC\,1705-D2, though diffuse emission at 8 \mm\ is prominent (compare
to Figure~\ref{figcap2}).  High-surface brightness PAH emission is
only detected in the eastern dust emission lobe NGC\,1705 D1; see also
the spectra of D1, D2, and the SSC in Figure~\ref{figcap4}.}
\label{figcap5}
\end{figure*}

Figure~\ref{figcap5} shows contours of a (continuum subtracted) IRS
spectral summation of the 11.3 \mm\ PAH emission, overlaid on the {\it
HST}/WFPC2 F380W and \halpha\ images. PAH emission is associated with
high-surface brightness \halpha\ emission in the eastern dust emission
complex (NGC\,1705 D1), suggesting that local radiation density is an
important parameter that affects the appearance of emission from these
molecules in the ISM \citep[but note the trend of decreasing PAH
emission strength with increasing radiation field density discussed
in][]{madden05conf}. However, radiation field strength alone seems to
be insufficient for the emergence of PAH emission: there is high
surface brightness \halpha\ emission associated with NGC\,1705 D2, but
no associated PAH emission is seen (see also the relative flux ratios
in Table~\ref{t1}). Previous studies have detected the 7.7 and 11.3
\mm\ PAH bands in the diffuse ISM of the Milky Way, unassociated with
regions of massive star formation and apparently excited by the local
interstellar radiation field \citep{lemke98}.  Other factors also
likely play important roles in the production and destruction of PAH
emission, including the local electron and gas densities
\citep{ruiterkamp02} and supernovae shocks \citep{ohalloran06}.

The extracted flux density in the 11.3 \mm\ PAH band in NGC\,1705 D1 is found
to be 2.4\pom0.8 mJy over a 0.745 \mm\ spectral region (see discussion in
\S~\ref{S2}).  Given the low S/N ratio of the associated 7.7 \mm\ PAH feature
in the IRS spectrum, its flux density can be estimated by assuming a uniform
surface brightness over the same aperture size; from Figure~\ref{figcap4}, a
peak value of $\sim$ 3.5 MJy\,sr$^{-1}$ is representative.  Scaled to the
10\arcsec\ diameter circular aperture used for IRAC flux extraction, this
corresponds to a flux density of $\sim$ 2.6 mJy.  If the continuum intensity
of $\sim$ 1 MJy\,sr$^{-1}$ seen between 9 and 10 \mm\ is representative of the
background under the 7.7 \mm\ feature, then this value will decrease to $\sim$
1.8 mJy.  These values can be compared with the residual emission in the IRAC
8 \mm\ band after continuum removal; the exemplary value of (IRAC 8 \mm) $-$
(0.232)$\cdot$(IRAC 3.5 \mm) used by \citet{helou04} yields a diffuse (stellar
subtracted) 8 \mm\ flux density of 3.0\pom0.5 mJy (though note that a single
scaling factor is likely not applicable to all galaxies).  Depending on the
underlying continuum levels, assumptions about the surface brightness
distribution of the PAH emission, and the scaling factor used to remove the
stellar continuum, it thus appears that most of the emission from region D1 in
the IRAC 8 \mm\ band arises from PAH emission.  This is also evident from
inspection of Figure~\ref{figcap4}; the 7.7 \mm\ PAH complex is clearly
detected in D1, but the spectrum of D2 is flat in this region.

These arguments imply that spatially variable PAH strengths compared
to the underlying dust continuum make dissecting PAH and warm dust
emission a complex issue. This is exemplified by region D2, which
shows high surface brightness \halpha, starlight-subtracted 8 \mm\ and
dust emission but no detectable PAH features.  Since the spectra of
photodissociation regions (PDRs) and of \HII\ regions are quite
different \citep[e.g.,][]{hollenbach97,genzel00}, the differences
between D1 and D2 may have to do with how much of a PDR is left in one
case vs the other.  IRS spectra should be used to verify the presence
and strength of PAH emission that may constitute an unknown fraction
of the detected diffuse emission at 8 \mm.

While the spectra presented in Figure~\ref{figcap4} are of relatively low S/N,
they offer interesting comparisons of the physical conditions in the three
regions probed.  NGC\,1705\,D1 shows pronounced [S~{\sc IV}] $\lambda$ 10.51
\mm\ emission, while this appears to be absent in D2 and the SSC.  The
[Ne~{\sc II}] 12.81 \mm\ line, which is blended with the 12.6 \mm\ PAH
feature, appears in the spectra of both D1 and D2.  The relative strengths of
[S~{\sc IV}] and [Ne~{\sc II}] in D1 compared to D2 argue for a stronger
radiation field in the former \citep[see discussion in][]{thornley00}.  The
low S/N of the spectra precludes a detailed treatment of the ionization
structure of the system.  The spectrum of the SSC shows a broad peak around 10
\mm\ that may be suggestive of emission from amorphous silicates, either ultra
small grains undergoing single-photon heating, or larger grains being heated
to steady temperatures of $\sim$ 250 K \citep[e.g.,][]{draine01}.

\subsection{The Empirical PAH Transition Metallicity}
\label{S3.3}

\citet{engelbracht05} find a pronounced change in the strength of PAH
emission as a function of metallicity in a sample of 34 star-forming
galaxies.  Above the level of 12+log(O/H) $=$ 8.2 ($\simeq$ 35\%
Z$_{\odot}$), PAH emission is detected via diffuse 8 \mm\ emission and
confirmed with IRS spectroscopy; at lower metallicities, PAH emission
is weak or absent in their sample.  NGC\,1705 has a nebular
metallicity equal to this transition point [12$+$log(O/H) $=$
8.21\pom0.05, measured at various locations throughout the central
disk (including both D1 and D2); see {Lee \& Skillman
2004}\nocite{lee04a}]; the variable PAH emission from different
regions of the galaxy provides an interesting test case for this
empirical relation.

\citet{engelbracht05} compare the 8/24 \mm\ ratio (R$_2$) to the 4.5/8
\mm\ ratio (after the 4.5 \mm\ image has had an estimate of the
underlying stellar continuum removed by scaling and subtracting the
IRAC 3.5 \mm\ image; R$_1$).  The 8/24 \mm\ ratio clearly separates
star forming galaxies according to both metallicity and the relative
strengths of stellar and dust emission: metal-rich galaxies
preferentially show high 8/24 \mm\ ratios and low ratios of stellar to
dust emission at 8 \mm, while metal-poor galaxies show opposite
trends.  These color separations are interpreted to be caused, to
first order, by the decreased emission strengths of PAH molecules in
the low-metallicity ISM.

Computing the diagnostic color ratios as described in
\citet{engelbracht05}, we find (R$_1$, R$_2$) values of
(0.05\pom0.005, 0.57\pom0.09), (0.14\pom0.009, 0.35\pom0.044), and
(0.14\pom0.02, 0.42\pom0.06) for NGC\,1705 D1, D2, and the total
galaxy, respectively, using the values in Table~\ref{t1}. Note,
however, that aperture corrections in the MIPS 24 \mm\ band can be
significant for the 10\arcsec\ diameter apertures used here (between
$\sim$ 20--50\%, depending on the extent of the source and the
contamination from neighboring emission regions).  Taken at face
value, each of the R$_2$ ratios is consistent with the average values
found in the PAH-emitting galaxies of the \citet{engelbracht05} study
(i.e., R$_2 >$ 0.2).  The R$_1$ ratio of the PAH emission region, D1,
falls exactly on the color boundary defined by \citet{engelbracht05};
the two other regions show higher R$_1$ ratios, consistent with values
found for systems with PAH non-detections.  If aperture corrections at
24 \mm\ are significant, the R$_2$ ratio will decrease for regions D1
and D2 (but not for the global measurements), and the
\citet{engelbracht05} diagnostic color ratios correctly separate the
regions D1 and D2 into PAH-emitting and PAH-deficient, respectively.

\subsection{Deviation from the Radio vs. Far-IR Correlation}
\label{S3.4}

In normal star-forming galaxies, there is a remarkably tight correlation
between the total IR luminosity and the strength of radio continuum emission
(see, e.g., {de~Jong \etal\ 1985}\nocite{dejong85}, {Helou \etal\
1985}\nocite{helou85}, {Condon 1992}\nocite{condon92}, and references
therein).  The tightness of the relation (scatter of less than 0.3 dex over
multiple orders of magnitude) is especially remarkable, given that the
physical processes that drive IR (re-radiation of absorbed photons by dust
grains) and radio emission (typically nonthermal synchrotron emission, though
a thermal free-free component can also contribute) are only indirectly
associated with one another. Indeed, \citet{bell03a} shows that the relation
holds for IR- and radio-deficient galaxies, but that these systems require a
complex interplay of star formation, dust content and magnetic fields to
produce the apparently linear relation in this regime.

We explore the relation between the global radio and IR luminosities
of NGC\,1705. The total IR (TIR) flux (derived by applying equation 4
from {Dale \& Helou 2002}\nocite{dale02}) and the radio continuum
luminosity appear to be related via a constant value in star-forming
galaxies [commonly denoted by the ``q'' parameter, where q $\propto$
log(S$_{\rm TIR}$/S$_{\rm RC}$); see, e.g., {Helou \etal\
1985}\nocite{helou85}, {Bell 2003}\nocite{bell03a}].
\citet{johnson03a} find a nondetection of NGC\,1705 in the radio
continuum at 6 cm, using data that are sufficiently sensitive to
detect the Galactic W\,49 star formation complex at the assumed
distance of 5.1 Mpc (rms $=$ 0.05 mJy\,bm$^{-1}$). Using the global
flux densities of NGC\,1705 in Table~\ref{t1}, we derive S$_{\rm TIR}$
$\simeq$ 8\,$\times$\,10$^{-14}$ W\,m$^{-2}$; assuming that ``q'' $=$
2.64 \citep{bell03a} and that the radio emission is a mix of thermal
and nonthermal processes (i.e., $\alpha = -$0.7, where \snu\ $\sim\
\nu^{\alpha}$) for the extrapolation from 20 to 6\,cm flux densities,
this would imply an expected radio continuum flux density of $\sim$ 2
mJy. This is more than an order of magnitude higher than the {Johnson
\etal\ (2003)}\nocite{johnson03a} upper limit.  Reducing the assumed
value of ``q'' (e.g., q $=$ 2.34; {Yun \etal\ 2001}\nocite{yun01},
though note that this value is derived using only information on the
SED between 42--122 \mm) will increase the expected flux density by
roughly 50\%; assuming a stronger nonthermal component in the
extrapolation from 1.4 to 4.8 GHz (e.g., $\alpha$ $= -$1.2, indicative
of pure synchrotron emission) reduces the expected value by $\sim$
50\%. Applying a monochromatic ``q'' parameter
\citep[e.g.,][]{murphy05} yields similar estimates of the expected
radio continuum flux density.  Regardless of the extrapolation, it is
clear that the radio continuum of NGC\,1705 is extremely weak compared
to the observed far-IR flux densities.

The detection, using these \spitzer\ data, of strong far-IR emission
shows that there is significant dispersion amongst individual galaxies
in the far-IR vs. radio correlation.  Indeed, \citet{bell03a} suggests
that this correlation is dependent on numerous factors and upon galaxy
type, and that both the IR and radio luminosities of dwarf galaxies
are suggested to significantly underestimate the true star formation
rate. It is important to note, by comparison with the internal scatter
in individual galaxies seen in the larger sample of \citet{murphy05},
that the tightness of the global radio-IR correlation may depend on
averaging over a significantly large number of emission regions.  In
dwarf galaxies, and starbursting dwarf galaxies in particular, the IR
and radio emission are typically dominated by one (or at most a few)
luminous star formation regions, allowing significant deviations from
the relations seen in more massive galaxies.  In larger starburst
galaxies, a deficiency of radio continuum emission compared to the
far-IR luminosity has been attributed to a very young ($\lsim$ few
Myr), ``nascent'' starburst that has occured after an extended period
($\gsim$ 100 Myr) of quiescence \citep[e.g., ][]{roussel03}.

In the case of NGC\,1705, which is IR-bright but radio-faint, the
aforementioned galactic wind may play an important role in the lack of radio
continuum emission. The young age of the SSC \citep[$\lsim$ 12
Myr;][]{vazquez04} implies that thermal emission associated with the UV
photons of massive stars should still be strong (see timescale arguments in
{Condon 1992}\nocite{condon92} and {Cannon \& Skillman
2004}\nocite{cannon04}).  Similarly, the ``post-starburst'' nature of the
system \citep{heckman97} and spatially resolved stellar photometry
\citep{tosi01,annibali03} have shown that most of the O stars associated with
the strong burst of star formation have exploded as SNe, implying a strong
synchrotron emission component.  If the burst has ``blown out'' of the disk
and is venting hot gas into the halo \citep[as evidenced by UV absorption line
spectroscopy and diffuse \halpha\ and x-ray gas in the halo;
see][]{hensler98,heckman01a,veilleux03}, the energetic particles that give
rise to radio continuum emission are likely escaping the disk \citep[see also
the discussion in][]{johnson03a}.

\section{Conclusions}
\label{S4}

We have presented observations of the nearby dwarf starburst galaxy
NGC\,1705 made with all three instruments on board the {\it Spitzer
Space Telescope}.  This is one of the few local dwarf galaxies known to
host a SSC, and the only such system in the \sings\ sample.  The
galaxy underwent a strong starburst some 15 Myr ago that produced the
SSC and massive star formation in the field; this starburst is driving
a multiphase outflow from the SSC region and this has dramatic
consequences for the multiwavelength properties of the galaxy.

The galaxy morphology changes significantly longward and shortward of
$\sim$ 5 \mm.  Optical and short-wavelength IRAC imaging shows an
underlying red stellar population, with the SSC dominating the
luminosity.  Longer-wavelength IRAC and MIPS imaging reveals warm dust
emission arising from two off-nuclear regions that dominate the far-IR
morphology of the system.  Stellar and warm dust emission from the SSC
is still prominent at 8 \mm, but the SED appears to flatten out toward
longer wavelengths, suggesting that most of the warm dust emission
arises from the off-nuclear regions. There is high surface brightness
\halpha\ emission associated with these dust emission peaks.  The
energy source for this off-nuclear nebular and dust emission, some
$\sim$ 250 pc from the SSC, is likely young massive stars in these
regions; some of these stars may be embedded, thus not contributing to
the observed \halpha\ luminosity.  The lack of extinction toward the
SSC \citep{heckman97} and the modest warm dust emission of the SSC in
the MIPS bands suggests that dust near the SSC may have been removed
by the multi-phase outflow or evacuated by the high UV flux from the
SSC.

We derive a total dust mass of $\sim$ (2.2\pom1.1)\,$\times$\,10$^5$
\msun.  This value represents the mean of the dust mass estimates from
using the SED models of \citet{dale01} and \citet{dale02} and the
models of \citet{li01,li02}.  A single-temperature blackbody function
is unable to simultaneously fit the observed 24, 70 and 160 \mm\ flux
densities.  Both sets of SED models predict a higher dust mass than a
single-temperature blackbody fit to the 70 and 160 \mm\ data points.
Comparing to the \HI\ mass implies a global dust-to-gas ratio
$\sim$ 2--4\,$\times$ lower than the Galactic average.

The strong diffuse emission in the IRAC 8 \mm\ band is dominated by warm dust
emission; IRS spectroscopy of the SSC and both off-nuclear sources shows 11.3
\mm\ PAH emission in the more luminous of the two dust emission peaks only (no
PAH emission is associated with the SSC). This serves as a prudent reminder
that IRAC diffuse 8 \mm\ emission traces both dust continuum and PAH emission,
and that spectroscopy is essential for a robust separation of the two
components.  The fact that variable PAH emission strengths are seen in a
galaxy with nebular metallicity equal to the empirical PAH threshold found by
\citet{engelbracht05} highlights the complexity in interpreting the nature of
diffuse 8 \mm\ emission.

Finally, we note that NGC\,1705 appears to deviate significantly from
the canonical far-IR vs. radio correlation, in the sense that strong
far-IR emission is present, but radio continuum emission remains
undetected at meaningful sensitivity levels. The galactic outflow may
play an important role in affecting the radio continuum emission; if
significant numbers of energetic particles are escaping into the halo,
the radio continuum emission will be significantly suppressed compared
to the total IR luminosity.  Dwarf starburst galaxies, typically
dominated by one massive star formation region, appear to be
susceptible to strong variations in this ratio; the tightness of the
radio-IR correlation may rely on a larger statistical sampling of star
formation regions throughout an individual galaxy.

\acknowledgements

The {\it Spitzer Space Telescope} Legacy Science Program ``The Spitzer
Nearby Galaxies Survey'' was made possible by NASA through contract
1224769 issued by JPL/Caltech under NASA contract 1407.  The authors
thank Henry Lee for helpful discussions, Gerhardt Meurer for providing
the \HI\ image used in our analysis, and the anonymous referee for an
exceptionally detailed report that helped to improve the manuscript.
This research has made use of the NASA/IPAC Extragalactic Database
(NED) which is operated by the Jet Propulsion Laboratory, California
Institute of Technology, under contract with the National Aeronautics
and Space Administration, and NASA's Astrophysics Data System.



\end{document}